\documentclass[aps,pre]{revtex4}

\usepackage{graphicx,psfrag}
\usepackage[hypertex]{hyperref}
\usepackage{amssymb}
\usepackage{color}

\renewcommand{\Re}{\mathrm{R}}
\renewcommand{\Im}{\mathrm{I}}

\newcommand{\D}{{\mathrm d}}
\newcommand{\E}{\mathrm{e}}

\newcommand{\average}[1]{\left<{#1}\right>}

\newcommand{\pxs}[1]{\frac{\partial^2 {#1}}{\partial x^2}}

\newcommand{\pqs}[1]{\frac{\partial^2 {#1}}{\partial Q^2}}

\newcommand{\p}[1]{\left({#1}\right)}
\newcommand{\pq}[1]{\left[{#1}\right]}

\newcommand{\derpart}[2]{\frac{\partial #1}{\partial #2}}

\begin{document}

\title{Probability density functions of work
and heat near the stochastic resonance of a colloidal particle}

\author{Alberto Imparato$^*$, Pierre Jop$^+$, Artyom Petrosyan $^+$\& Sergio Ciliberto $^+$}
\affiliation{ $^*$   Department of Physics and Astronomy, University of Aarhus, DK-8000 Aarhus C, Denmark
 \\ $^+$Universit\'e de
Lyon, Laboratoire de Physique de l'\'Ecole Normale Sup\'erieure de
Lyon, CNRS UMR 5276, , 46 all\'ee d'Italie, 69364 Lyon cedex 7,
France.}

\date{\today}


\begin{abstract}
We study experimentally and theoretically the
probability
 density functions  of the injected and dissipated  energy in a system of
a colloidal particle trapped in a double well potential
periodically  modulated by an external perturbation. The work done
by the external force and the dissipated energy are measured close
to the stochastic resonance where  the injected power is maximum.
We show a good agreement between the  probability density functions exactly computed from a
Langevin dynamics and the measured ones.  The probability density function of the work done on the particle satisfies the
fluctuation theorem.
\end{abstract}

\maketitle

\section{Introduction}

The study of fluctuations of the injected and dissipated power in
a system driven out of equilibrium by an external force is
nowadays a widely studied problem which is not yet completely
understood. This is a very important and general issue within the
context of Fluctuation Theorems (FT)  which constitute extremely
useful relations for characterizing the probabilities of observing
entropy production or consumption in out of equilibrium systems.
These relations were first observed in the simulations of a
sheared fluid\cite{evans} and later proved both for chaotic
dynamical systems \cite{gallavotti95} and for stochastic
dynamics\cite{kurchan98}. These works lead to different
formulations which find powerful applications for measuring
free-energy difference in biology (see e.g. \cite{ritort06} for a
review). The hypothesis and the extensions of fluctuation
theorems\cite{Cohen}
 have been tested in various
 experimental systems such as colloidal
 particles\cite{blickle06,wang02,al_pre07}, mechanical
 oscillators\cite{joubaud07}, electric circuits\cite{garnier} and optically driven
single two-level systems\cite{schuler05}.  The effect of
anharmonic potential on the motion of a colloidal particle has
been  tested by Blickle et al
 \cite{blickle06,seifert07,schuler05}. In
a recent experiment \cite{jop} it has been shown that FT holds for
a  colloidal particle confined in a double well potential and
driven out of equilibrium near the stochastic resonance (see next
section). The purpose of this article is to compare the probability density function (PDF) for
work and heat measured in this experiment  with those analytically
computed from a non-linear Langevin dynamics. We also discuss the
difference between the PDF obtained from a phase average of the
driving and those obtained from a fixed phase only. Such a precise
comparison between theory and experiment  in a double well
potential driven out of equilibrium has never been done before.
There is only
 a numerical study, which has explored the distributions
of the dissipated heat and of the work in a Langevin dynamics near
the stochastic resonance \cite{saikia07,sahoo07}. The paper is
organized as follow. The properties of the stochastic resonance
are recalled in the next section. The experimental set-up is
described in section 3. The analytical  PDF are derived  in
section 4. The comparison between the theoretical predictions and
the experimental measurements is done in section 5. Finally we
conclude in section 6.

\section{Stochastic resonance}
A colloidal particle, confined in a double well potential, hops
between the two wells at a rate $r_k$, named the Kramers' rate,
which is determined by the height $\delta U$ of the energy barrier
between the two wells, specifically $r_k=\tau_0^{-1}
\exp({-{\delta U \over k_BT}})$, where $\tau_0$ is a
characteristic time,  $k_B$ the Boltzmann constant and $T$ the
heat bath temperature \cite{libchaber}. When the double well
potential $U$ is modulated by an external periodic perturbation
whose frequency is close to $r_k$ the system presents the
stochastic resonance phenomenon\cite{Benzi}, i.e. the hops of the
particle between the two wells synchronize with the external
forcing. The stochastic resonance has been widely studied in many
different systems and it has been shown to be a bona fide
resonance looking at the resident time\cite{Benzi,gammaitoni95},
the Fourier transform of the signal for different noise
intensity\cite{babic04}. Numerically, the stochastic resonance has
been characterized by computing the injected work done by the
external agent as a function of noise and frequency
\cite{iwai01,dan05}. In a recent experiment \cite{jop} some of us have
studied  experimentally the  Steady State  Fluctuation Theorem (SSFT) in a system composed by  a Brownian
particle trapped in a double well potential periodically modulated
by an external driving force. We have measured  the energy
injected into the system by the sinusoidal perturbation and we
have analyzed the distributions of work and heat fluctuations. We
find that although the dynamics of the system is strongly
non-linear the SSFT holds for the work integrated on time
intervals which are only a few periods of the driving force. In
this paper we will compare this measured PDF with those derived
analytically from a Langevin dynamics in which the exact potential
of the experiment has been used.

\section{Experimental Set-up}

The experimental setup is composed by a custom built inverted
optical tweezers made of an oil-immersion objective (63$\times$,
N.A.=1.3) which focuses a laser beam (wavelength
$\lambda=1064$~nm) to the diffraction limit
  for trapping glass beads ($2~\mu$m in diameter). The silica beads are dispersed
 in bidistilled water in very small concentration. The suspension is introduced
 in the sample chamber of dimensions $0.25\times10\times10$~mm$^3$,
 then a single bead is trapped and moved away from others.\\
The position of the bead is tracked using a fast-camera with the
resolution of 108 nm/pixels which gives after treatment the
position of the bead with an accuracy better than 20~nm.
 The trajectories of the bead are sampled at
50~Hz. The position of the trap can be easily displaced on the
focal plane of the objective by deflecting the laser beam using an
acousto-optic deflector (AOD). To construct the double well
potential the laser is focused alternatively at two different
positions at a rate of 5 kHz. The residence times $\tau_i$ (with
$i=1,2$) of the laser in each of the two positions determine the
mean trapping strength felt by the trapped particle. Indeed if
$\tau_1=\tau_2=100\mu s$ the typical diffusion length of the bead
during this period is only  5~nm. As a consequence the bead feels
an  average   double-well potential:
\begin{equation}
U_0(x)=ax^4-bx^2-dx \ ,
\end{equation}
 where $a$, $b$ and $d$ are
determined by the laser intensity and by the distance of the two
focal points. In our experiment the distance between the two spots
is $1.45~\mu$m, which produces a trap whose minima are at
$x_{min}=\pm$0.45 $\mu$m. The total intensity of the laser is
$58$~mW on the  focal plane which corresponds to an inter-well
barrier  energy $\delta U_0=1.8~k_BT$. Starting from the static
symmetric double-trap, ($\tau_1=\tau_2$) we modulate the depth of
the wells at low frequency by modulating the residence times
($\tau_i$) during which the spot remains in each position. We
 keep the total intensity of the laser constant
  in order to produce a more stable potential.
 The modulation of the average intensity is harmonic at frequency $f$ and its  amplitude
 $(\tau_2-\tau_1)/(\tau_2+\tau_1)$,
  is $0.7~\%$ of the average intensity in the static symmetric case.
 Thus the potential felt by the bead has the following profile in the axial direction:
\begin{equation}
U(x,t)=U_0(x)+U_p(x,t)=U_0+c\ x \ \sin(2 \pi f t),
\label{eq_U}
\end{equation}
with $ax_{min}^4= 1.8\  k_BT$, $bx_{min}^2=3.6 \ k_BT$,
$d|x_{min}|=0.44\ k_BT$ and $c|x_{min}|=0.81~k_BT$. The amplitude
of the time dependent perturbation is synchronously acquired with
the bead trajectory. The parameters given here are average
parameters since the coefficients $a$,
 $b$ and $c$ ,obtained from fitted steady distributions at given phases,
 vary with the phase ($\delta a/a\approx10\%$, $\delta b/b\approx\delta c/c\approx5\%$).

\begin{figure}[htbp]
\begin{center}
\includegraphics[width=8cm]{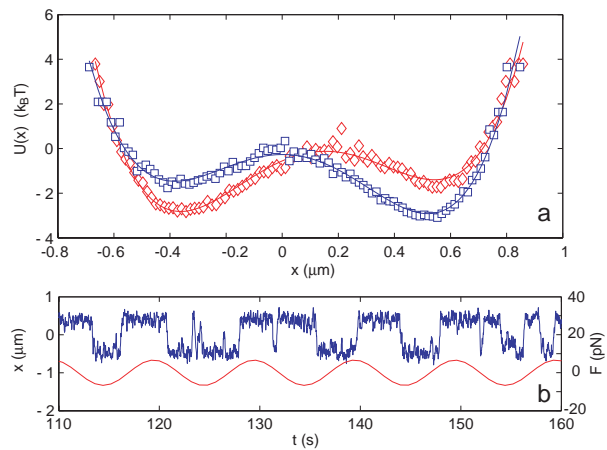}
\caption{a) The perturbed potential at $t=\frac{1}{4 f}$ and half
a forcing period later. b) Example of trajectory of the glass bead
and the corresponding perturbation at $f=0.1$ Hz.}
\label{fig:artforcingex}
\end{center}
\end{figure}

An example of the measured potential at $t=\frac{1}{4 f}$ and at
$t=\frac{3}{4 f}$ is shown on the Fig.~\ref{fig:artforcingex}a).
 The figure is obtained by measuring the two steady
state probability distribution function $P(x)$ of $x$,
(corresponding to $U_0+c$ and $U_o-c$ respectively)
 and by taking $U(x)=-\ln\left[P(x)\right]$.

\subsection{The equation of motion.}
The $x$ position of the particle can be  described by a Langevin
equation:
\begin{equation}
    \gamma\dot x=-\frac{\partial U}{\partial x}+\xi,
\label{lang_eq}
\end{equation}
with $\gamma=1.61 \ 10^{-8}$ N s m$^{-1}$ the friction coefficient
and $\xi$ the stochastic force. The natural Kramers' rate ($c=0$)
for the particle is $r_k=0.3$ Hz at $T=300$ K. When $c\ne0$  the
particle can experience a stochastic resonance when the forcing
frequency is close to the Kramers' rate. An example of the
sinusoidal force with the corresponding position are shown on the
figure \ref{fig:artforcingex}b).

\subsection{The work and the heat.}

Since the synchronization is not perfect, sometimes the particle
receives energy from the perturbation, sometimes the bead moves
against the perturbation leading to a negative work on the system.

In the following, all energies are normalized by $k_BT$. From the
trajectories, we compute  the stochastic $W_s$ and the classical
$W_{cl}$ works done by the perturbation on the system and the heat
$Q$ exchanged with the bath. These three quantities  are defined
by the following equations as  in ref.\cite{sekimoto96}:
\begin{eqnarray}
    W_s(t_0,t_f)=\int^{t_0+t_f}_{t_0}{dt\frac{\partial U(x,t)}{\partial t}}&&  \nonumber \\
 W_{cl}(t_0,t_f)=-\int^{t_0+t_f}_{t_0}{dt\dot x\frac{\partial
U_p(x,t)}{\partial x}}&&
    \label{eq:integr}\\
    Q(t_0,t_f)=-\int^{t_0+t_f}_{t_0}{dx\frac{\partial U(x,t)}{\partial x}}&& \nonumber
\end{eqnarray}
where in this case $t_f={n \over f}$ is a multiple of the forcing
period. We use both $W_s$ and $W_{cl}$ because they give
complementary information on the fluctuations of the energy
injected by the external perturbation into the system (see ref.
\cite{Taniguchi} and reference therein for a discussion on this
point). For example, as discussed in ref.\cite{jop}  $W_s/T$ is
the total entropy production rate in this specific case
\cite{seifert07b}.
The heat and the work, defined in eq.\ref{eq:integr}, are related
 through the first principle of thermodynamics: $Q=-\Delta U+W_s$,
  where $\Delta U=U(x(t_f+t_0),t_0+t_f)-U(x(t_0),t_0)$, whereas the two works
  are related by a boundary term $W_{cl}=-\Delta U_p+W_s$,
  where $\Delta U_p=U_p(x(t_f+t_0),t_f+t_0)-U_p(x(t_0),t_0)$.
 Since the characteristic time evolution of the perturbation is small compared to the fluctuation of
 position and due to the harmonic form of the perturbation,
 the integrals are computed as follows:
\begin{eqnarray}
    W_s(t_0,t_f)&=&\omega \ c \ \delta t \ \sum_{i=1}^{t_f/\delta t}x(i)\cos(\omega (t_0+t_i))  \nonumber\\
    W_{cl}(t_0,t_f)&=&-\Delta U_p+W_s
    \label{eq:wqcompute}\\
    Q(t_0,t_f)&=&-\Delta U+W_s  \nonumber
\end{eqnarray}
where $\delta t$ is the sampling time. We checked that the direct
computing of integrals of $Q$ and $W_{cl}$ gives the same results.
It is important to stress that $t_0$ can either take any value
(as it has been done in ref.\cite{jop}) or be a multiple of $1/f$:
 the fluctuations in the two
cases exhibit different PDFs, as we will see in the next sections.
To compute the works and heat from experimental data for a given duration $t_f$,
we thus divide a single trajectory into different segments starting either with a fixed phase, or with different phases, before averaging the results over the whole trajectory, and then over different runs.
In ref.\cite{jop}  the average work received over one period has
been measured  for different frequencies ($t_f={1\over f}$ in
eq.~\ref{eq:integr}). Each trajectory is here recorded during
3200~s in different consecutive runs, which corresponds to 160 up
to 7500 forcing periods, for the range of frequencies explored. It
has been found that the maximum injected energy is  around the
frequency $f \approx 0.1$~Hz, which is comparable with half of the
Kramers' rate of the fixed potential $r_K=0.3$~Hz. This maximum of
transferred energy shows that the stochastic resonance for a
Brownian particle is a bona fide resonance, as it was previously
shown in experiments using resident time distributions
\cite{gammaitoni95,schmitt06} or directly in simulations
\cite{iwai01,dan05}.
 It is worth noting that the average values of work in the  case of a periodic forcing do not
 dependent on their definitions: only the boundary terms, which
 vanish in average with time, are different \cite{jop}. As the
 difference between $W_{cl}$ and $W_s$ has been discussed in
 ref.\cite{jop} we will focus here only on $W_s$ that, in order to simplify
 the notation,  will be indicated by $W$.


\section{Equations for the W and Q PDFs}

In this section we discuss the equation governing the time evolution of
the work and heat PDFs.
For a stochastic process described by eq.~(\ref{lang_eq}) the Fokker-Planck equation reads
\begin{equation}
\partial_t p(x,t)=\Gamma \derpart{}{x}\pq{U'(x,t) p}+k_B T \Gamma \frac{\partial^2 p}{\partial x^2},
\label{fp_eq}
\end{equation}
where $p(x,t)$ is the PDF associated to the coordinate $x$, and
$\Gamma=1/\gamma$. Here and in the following, the prime denotes derivative with respect to $x$.

Let us consider the joint probability distribution function of the
position and of the stochastic work $\phi(x,W,t)$: in ref.~\cite{al_w} it has been shown that the time evolution of such
a function is governed by  the partial differential equation
\begin{equation}
\partial_t \phi(x,W,t)=\Gamma \derpart{}{x}\pq{U'(x,t) \phi}+k_B T \Gamma \frac{\partial^2 \phi}{\partial x^2}-\derpart{U}{t}  \derpart{\phi}{W},
\label{eq_phi}
\end{equation}
with the starting condition $\phi(x,W,t_0)=p(x,t_0)\delta(W)$.
The
unconstrained probability distribution of the work is given by
\begin{equation}
\Phi(W,t)=\int d x \phi(x,W,t). \label{uncW}
\end{equation}
By introducing the Fourier transform
\begin{equation}
\psi(x,\lambda,t)=\int d W \E^{-\lambda W} \phi(x,W,t),
\label{trasf_phi}
\end{equation}
eq.~(\ref{eq_phi}) becomes
\begin{equation}
\partial_t \psi(x,\lambda,t)=\Gamma \derpart{}{x}\pq{U'(x,t) \psi}+k_B T\Gamma \frac{\partial^2 \psi}{\partial x^2}-\lambda \derpart{U}{t} \psi,
\label{eq_psi}
\end{equation}
with the starting condition $\psi(x,\lambda,t_0)=p(x,t_0)$.
Note that, by using the Fourier transform definition eq.~(\ref{trasf_phi}),
the wavenumber $\lambda$ associated to $W$ is a purely imaginary number, $\lambda=i |\lambda|$.

Let us now consider the  joint probability distribution $\varphi(x,Q,t)$ of the position
$x$ and the heat $Q$  exchanged by the brownian particle
whose motion is described by eq.~(\ref{lang_eq}).
The Fokker-Planck-like equation, governing the time evolution of such a function, reads
\cite{al_pre07,SpSe}
\begin{eqnarray}
\partial_t \varphi(x,Q,t)&=& \partial_x \p{ \Gamma U' \varphi}-\partial_Q (\Gamma U'^2 \varphi)
-\partial_x\pq{(\Gamma U' k_B T) \partial_Q \varphi}-\partial_Q  \pq{(\Gamma U' k_B T) \partial_x \varphi}\nonumber \\
&& +\Gamma k_B T \pxs{\varphi}+ \Gamma k_B T U'^2\pqs{\varphi} ,
\label{var_phi_eq}
\end{eqnarray}
with the starting condition $\varphi(x,Q,t_0)=p(x,t_0)\delta(Q)$.
Note that we use here the opposite sign convention for $Q$, eq. (\ref{eq:wqcompute}),
with respect to that adopted in ref.~\cite{al_pre07}.

The unconstrained probability distribution functions of the heat reads
\begin{equation}
\mathit{\Phi}(Q,t)=\int \D x\, \varphi(x,Q,t).
\end{equation}

Let $\chi(x,\lambda,t)$ be the Fourier transform of
$\varphi(x,Q,t)$, as given by
\begin{equation}
\chi(x,\lambda,t)=\int \D Q \,\E^{-\lambda Q} \varphi(x,Q,t),
\label{chi_def}
\end{equation}
eq.~(\ref{var_phi_eq}) becomes
\begin{equation}
\partial_t \chi(x,\lambda,t)=\Gamma k_B T \pxs{\chi}+\partial_x \p{ \Gamma U' \chi}-\lambda \Gamma U'^2 \chi- \lambda \partial_x\pq{(\Gamma U' k_B T) \chi}-\lambda (\Gamma U' k_B T) \partial_x \chi+  \lambda^2 \Gamma k_B T U'^2\chi,
\label{gen_q}
\end{equation}
with the starting condition $\chi(x,\lambda,t_0)=p(x,t_0)$.
In ref.~\cite{al_pre07}, it has been shown that by introducing
 the function $g(x,\lambda,t)$ defined by
\begin{equation}
\chi(x,\lambda,t)=g(x,\lambda,t) \exp\pq{-\frac {\beta -2 \lambda} 2 U(x,t)},
\label{def_g}
\end{equation}
equation (\ref{gen_q}) simplifies, as one gets rid of the first order derivatives with respect to $x$, and it becomes
\begin{equation}
\partial_t g=\Gamma k_B T \pxs{g}-\Gamma \beta  \frac{U'^2}{4} g +\frac \Gamma 2 U''g +\frac {\beta- 2 \lambda}2 g \, \partial_t U,
\label{eq_g}
\end{equation}
with the starting condition as given by
$g(x,\lambda,t_0)=p(x,t_0)\exp[(\beta -2 \lambda)U(x,t_0)/2 ]$.

We now show that a similar simplification can be obtained for eq.~(\ref{eq_psi}):
let $h(x,\lambda,t)$ be defined as
\begin{equation}
\psi(x,\lambda,t)=h(x,\lambda,t)\exp[-\beta U(x,t)/2],
\label{def_h}
\end{equation}
substituting $h(x,\lambda,t)$ into eq.~(\ref{eq_psi}), we obtain the following equation for  $h(x,\lambda,t)$:
\begin{equation}
\partial_t h=\Gamma k_B T \pxs{h}-\Gamma \beta  \frac{U'^2}{4} h +\frac \Gamma 2 U''h +\frac {\beta -2 \lambda}2 h\,  \partial_t U.
\label{eq_h}
\end{equation}
The last equation is identical to
 (\ref{eq_g}), however its starting condition reads
$ h(x,\lambda,t_0)=p(x,t_0)\exp[\beta U(x,t_0)/2 ]$, which is
different from the starting condition of eq.~(\ref{eq_g}).

As $\lambda$ is an imaginary number, one can split eq.~(\ref{eq_g}) and
eq.~(\ref{eq_h}), in a set of two equations for the real and
imaginary part. For example, eq.~(\ref{eq_h}) becomes
\begin{eqnarray}
\partial_t h_\Re&=&\Gamma k_B T \pxs{h_\Re}-\Gamma \beta  \frac{U'^2}{4} h_\Re +\frac \Gamma 2 U''h_\Re +\p{\frac \beta 2 h_\Re +|\lambda| h_\Im}   \partial_t U, \label{h_r}\\
\partial_t h_\Im&=&\Gamma k_B T \pxs{h_\Im}-\Gamma \beta  \frac{U'^2}{4} h_\Im +\frac \Gamma 2 U''h_\Im +\p{\frac \beta 2 h_\Im -|\lambda| h_\Re}   \partial_t U, \label{h_i}
\end{eqnarray}
where $h_\Re$ and $h_\Im$ are the real and the imaginary part of $h$, respectively.

Equations (\ref{h_r}) and (\ref{h_i}), and the analogous equations for $g_\Re$ and $g_\Im$, can be solved numerically for any value of $|\lambda|$, and for any choice of the initial condition $p(x,t_0)$ using, e.g., MATHEMATICA \cite{mat_book}. The functions $\psi(x,\lambda,t)$ and $\chi(x,\lambda,t)$, can be thus obtained by using equations~(\ref{def_h}) and (\ref{def_g}).
The target functions $\phi(x,W,t)$ and $\varphi(x,Q,t)$ can then be obtained
by taking the Fourier inverse transform, i.e. by inverting eqs.~(\ref{trasf_phi}) and
 (\ref{chi_def}), respectively. Also in this case the computation can be performed numerically.

\section{Comparison with experiments}

In this section, we compare the results for the work and heat
PDFs, as obtained by solving the differential equations introduced
in the previous section, with the experimental outcomes. We take
the external driving frequency to be equal to $f=0.25$ Hz, to have
a good statistic, by allowing the observation of the system over a
sufficient number of periods. Such a value is  close to the
natural Kramers' rate, which ensures that the system is in the
stochastic resonance regime.

\begin{figure}[h]
\center
\includegraphics[width=8cm]{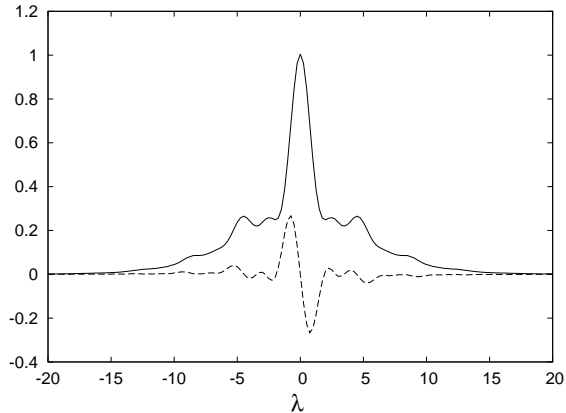}
\caption{Unconstrained Fourier transform of the work PDF as a function of $\lambda$, as obtained by numerical solution of eq.~(\ref{eq_psi}), with $t_0=50/f$ and
$t_1=t_0+1/f$. Full line $\Psi_\Re(\lambda,t_1)$, dashed line $\Psi_\Im(\lambda,t_1)$.}
\label{psi_lambda}
\end{figure}

\begin{figure}[h]
\includegraphics[width=8cm]{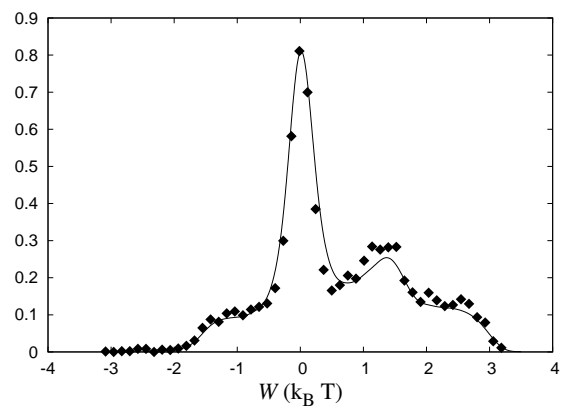}
\caption{Diamonds: experimental probability distribution of the work done on the
particle, along a single period, and with a single value of the initial phase. Full line: PDF of the work as
obtained by numerical solution of
eqs.~(\ref{eq_phi})-(\ref{eq_psi}), with $t_0=50/f$ and
$t_1=t_0+1/f$.} \label{figW0}
\end{figure}

We first consider the distribution of the work over a single
period of the external potential (\ref{eq_U}). The unconstrained
probability distribution of the work $\Phi(W,t)$, is obtained as
follows. In order to solve eq.~(\ref{eq_psi}), we solve
numerically eq.~(\ref{fp_eq}) up to the time $t_0=50/f\gg1/f$, so
as to ensure that the solution $p(x,t_0)$ of eq.~(\ref{fp_eq})
represents the steady state distribution of the position of the
particle. Such a solution is used as a starting condition for
eq.~(\ref{eq_psi}), which is solved numerically up to time
$t_1=t_0+1/f$, i.e. along a single period, and for different
values of $\lambda$, so as to obtain $\psi(x,\lambda,t_1)$. The
resulting unconstrained Fourier transform of the work PDF, defined
as $\Psi(\lambda,t_1)=\int \D x\,\psi(x,\lambda,t_1)$, is plotted
in fig.~\ref{psi_lambda} .
 As discussed in the previous section, by inverting
eq.~(\ref{trasf_phi}) and exploiting eq.~(\ref{uncW}), we finally
obtain $\Phi(W,t_1)$. In fig.~\ref{figW0}, the experimental histogram of the
work done on the particle along a single period $1/f$ is plotted;
in the same figure the unconstrained probability distribution of
the work $\Phi(W,t_1)$ is also plotted. We find a good agreement
between the experimental distribution of the work and the expected
PDF. It is worth noting first that we start sampling $W$ only few periods $1/f$ after the beginning of each experiment and second that the experimental results are averaged over different values of $t_0$ in order to improve the statistics (either with a fixed phase or over different phases).

\begin{figure}[h]
\center
\includegraphics[width=8cm]{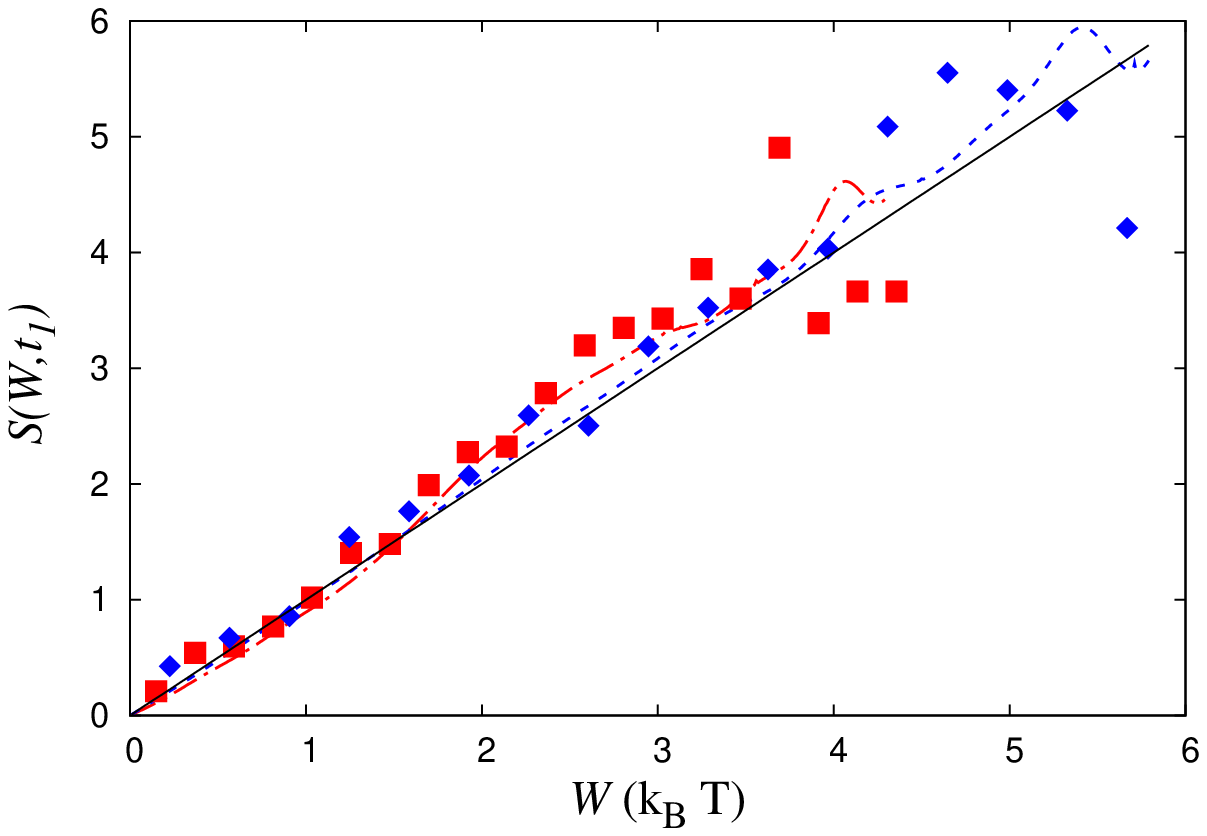}
\includegraphics[width=8cm]{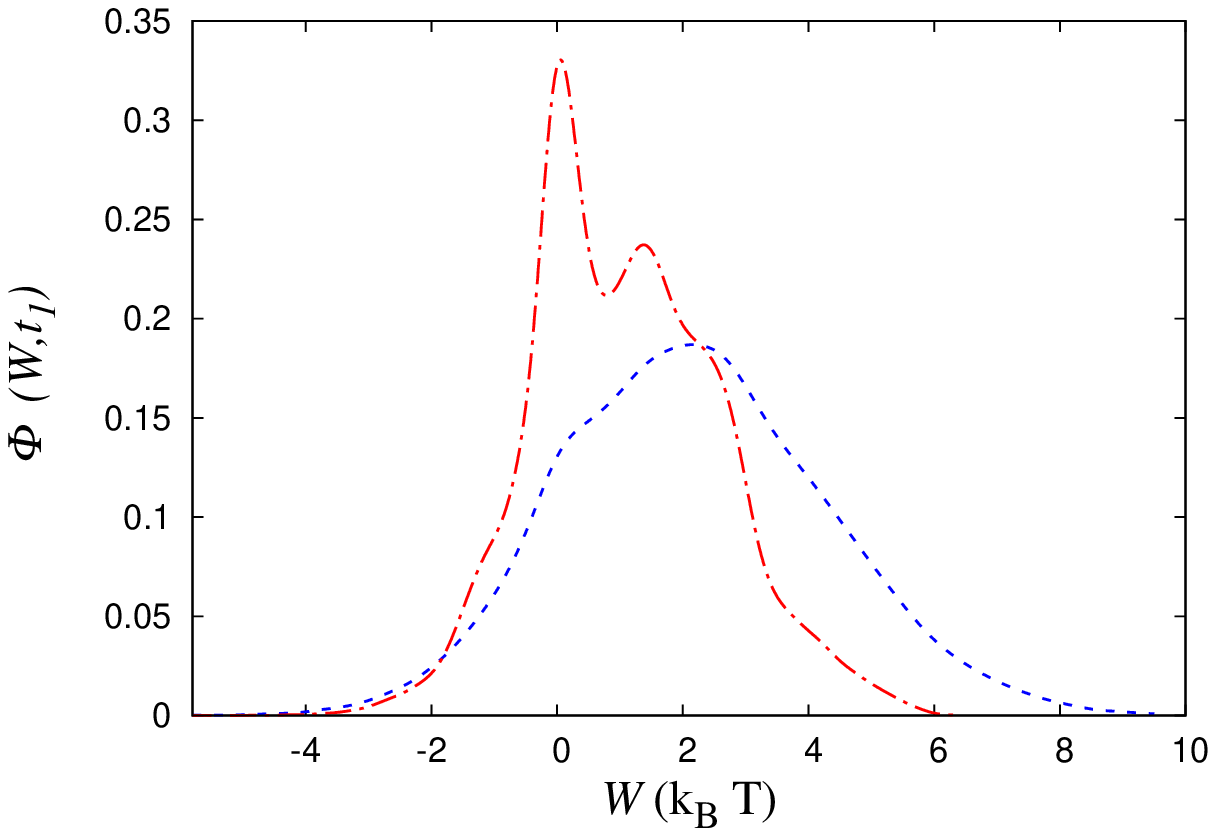}
\caption{Right panel: Symmetry function for the work PDF $S(W,t_1)$, eq.~(\ref{sw_def}), for two different values of $t_1$. The symbols corresponds to the experimental
data, the lines to the numerical solution of
eqs.~(\ref{eq_phi})-(\ref{eq_psi}) with $t_0=50/f$ .
Red boxes and red dashed-dotted line: symmetry function obtained with $t_1=t_0+2/f$. Blue diamonds and blue dashed line: symmetry function obtained with $t_1=t_0+4/f$. The full line corresponds to the expected behaviour
$S(W,t_1)=W$.  Left panel: PDF of the work for  $t_1=t_0+2/f$ (red dashed-dotted line), and  $t_1=t_0+4/f$ (blue dashed line).} \label{ft_fig}
\end{figure}

We now check that the  work PDF satisfies the SSFT, which we expect to be true for any time $t_1$ which is an integer multiple of the period $1/f$ as discussed in \cite{seifert07,seifert05}.  Given the periodicity of the external driving, the stochastic work here
considered corresponds to the total entropy production, as defined by Seifert in \cite{seifert07,seifert05}.
In figure~\ref{ft_fig}, we plot the symmetry function for the work PDF,
defined as
\begin{equation}
S(W,t_1)=\log\pq{\Phi(W,t_1)/\Phi(-W,t_1)},
\label{sw_def}
\end{equation}
together with the corresponding work PDF, for two values of the final time $t_1$.
 The fluctuation theorem
states that the simmetry function has to satisfy the equality
$S(W,t_1)=W$, for any value of $W$. Inspection of the figure
indicates that the fluctuation theorem is satisfied even at short
times $t_1$, the small deviations for the largest value of $W$
being due to numerical and experimental difficulties in the
sampling of the tails of the work PDF. More details
on SSFT have been already discussed in ref.~\cite{jop}.

\begin{figure}[h]
\includegraphics[width=8cm]{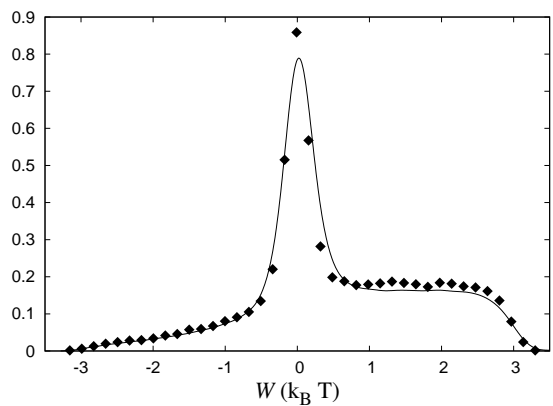}
\caption{Diamonds: experimental probability distribution of the work done on the
particle averaged over different initial phases. Full line, PDF of the work $\overline \Phi(W)$ as obtained by
numerical solution of eqs.~(\ref{eq_phi})-(\ref{eq_psi}), averaged
over $N=20$ different phases, see text.} \label{figW_ave}
\end{figure}

We now consider the work distribution averaged over several
phases. In order to obtain the probability distribution of the
work,  we solve eq.~(\ref{fp_eq}) up to the time $t_0(k)=(50+k/N)/
f$, where $N=20$ and $k=0,\dots N-1$, i.e. we consider $N$ initial
conditions with a phase difference $1/(N f)$. For each value of
$k$ we solve eq.~(\ref{eq_psi}), up to time $t_1(k)=t_0(k)+1/f$,
so as to obtain $\psi(x,\lambda,t_1(k)|p(x,t_0(k))$, i.e. the
Fourier transform of the work PDF with the starting condition
$p(x,t_0(k))$, and by antitransforming, we obtain
$\phi(x,W,t_1(k)|p(x,t_0(k))$. Finally the average work PDF is
obtained as  $\overline \Phi(W)=1/N \sum_k\int d x
\phi(x,\lambda,t_1(k)|p(x,t_0(k))$. In fig.~\ref{figW_ave}, this
distribution is compared with the experimental outcomes. We find a
good agreement
between the expected PDF and the histogram of the work.

\begin{figure}[h]
\center
\includegraphics[width=8cm]{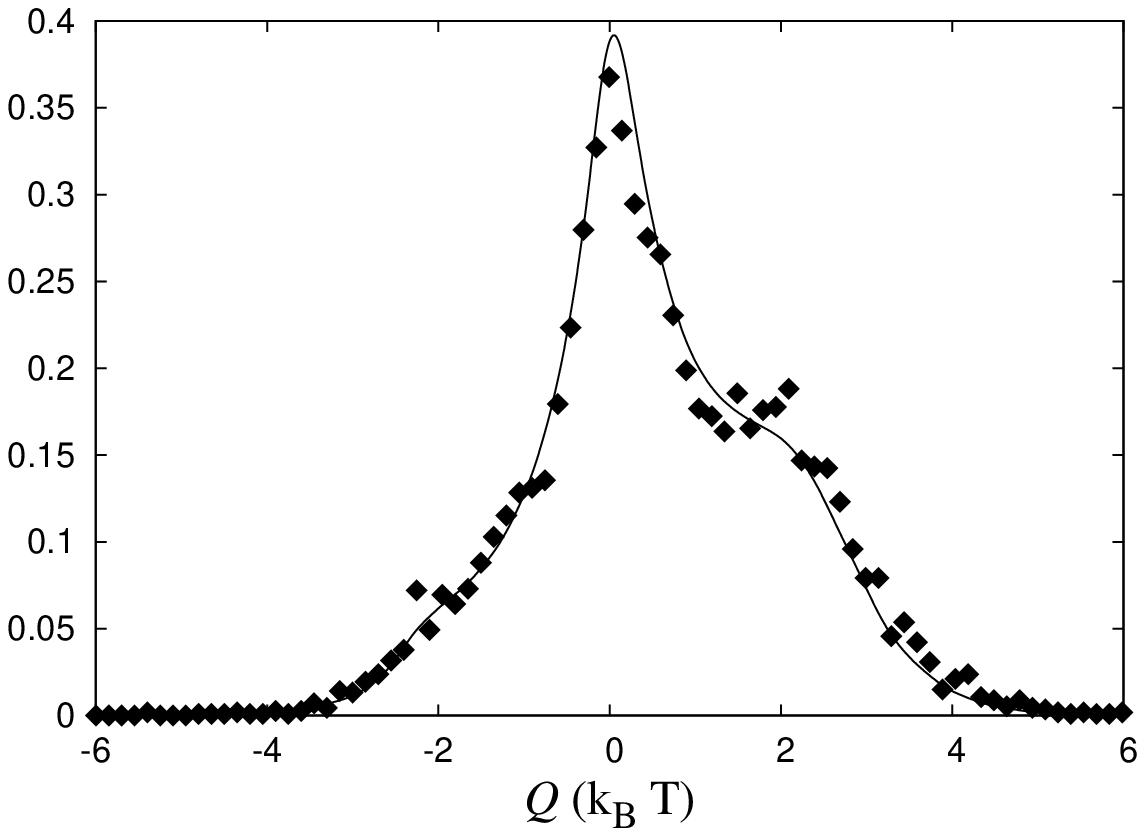}
\caption{Diamonds: experimental probability distribution of the heat exchanged by the particle, along a single period, and with a single value of the initial phase. Full line,  $\mathit \Phi(Q,t_1)$ as
obtained by numerical solution of
eqs.~(\ref{var_phi_eq})-(\ref{gen_q}), with $t_0=50/f$ and $t_1=t_0+1/f$.}
\label{single_Q}
\end{figure}

\begin{figure}[h]
\includegraphics[width=8cm]{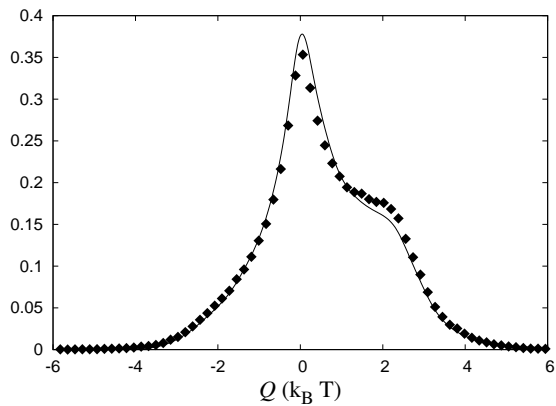}
\caption{Diamonds: experimental probability distribution of the heat exchanged by the
particle with the bath averaged over different initial phases. Full line, heat PDF $\mathit{\overline \Phi}(Q)$ as
obtained by numerical solution of
eqs.~(\ref{var_phi_eq})-(\ref{gen_q}), averaged over $N=20$ different
phases, see text.} \label{figQ_ave}
\end{figure}

Similarly to what has been done for the work, we now consider the
PDF of the heat $\mathit{\Phi}(Q)$, both with a single value of the
initial phase, and averaged over several initial conditions.
Also in this case, we use the function $p(x,t_0)$, solution of eq.~(\ref{fp_eq}),
as a starting condition for the numerical integration of eq.~(\ref{gen_q}),
then by antitransforming the function
$\chi(x,\lambda,t_1)$, we obtain  $\varphi(x,Q,t_1)$ and finally $\mathit{\Phi}(Q,t_1)$,
as discussed in the previous section.
The results for the case of  a single value of the initial phase is plotted
 in fig.~\ref{single_Q}, where we show a nice agreement with the experimental data.
In order to obtain the heat PDF averaged over several initial phases, we
solve numerically eq.~(\ref{gen_q}) in the time intervals
$[t_0(k),t_1(k)]$, where $t_0(k)$, and $t_1(k)$ have been defined above. Then by antitransforming the function
$\chi(x,\lambda,t_1(k))$, we obtain  $\varphi(x,Q,t_1|p(x,t_0(k))$,
and finally the unconstrained PDF of $Q$ defined as
$\mathit{\overline \Phi}(Q)=1/N \sum_k\int d x \varphi(x,Q,t_1|p(x,t_0(k))$.
In fig.~\ref{figQ_ave} we compare this expected PDF with the
experimental outcomes, the agreement between the numerical PDF and the experimental data is good also in this case.

\section{Conclusions}

In conclusion, we have experimentally and theoretically
investigated the power injected in a bistable colloidal system by
an external oscillating force.

 We have compared the PDFs of the stochastic work and of the
heat measured in the experiment with those obtained theoretically.
They exhibit large tails toward negative values and their shape are
non-gaussian close to the resonance when averaging over a single
period.

In refs.~\cite{al_pre07}, it has been shown that if the external
potential is quadratic, an analytical gaussian solution can be
obtained for  equation (\ref{eq_phi}) at any value of $t$ and for
eq.~(\ref{var_phi_eq})  in the long time limit. Here we have shown
that the same equations can be solved numerically, also at finite
time and for a more complex potentials, and they lead predictions
on the system energy exchange that can be verified experimentally.
We have shown that the Fourier transform of the heat and work PDFs
obey the same differential  equations (\ref{eq_g}) and
(\ref{eq_h}), and the only difference lies in the initial
conditions. This ensures $\Phi(W)$ and $\mathit \Phi(Q)$ to be
different at short time, while one has $\Phi(W)=\mathit \Phi(Q)$
in the long time regime, as one would expect, since
$\average{U}=0$ for long times.

Thus, our results suggest that the set of equations (\ref{eq_phi})
and (\ref{var_phi_eq}) represents an useful and general tool to
investigate the energy balance of microscopic systems in arbitrary
potentials.

Finally, we find that the work PDF satisfy the fluctuation theorem, even at
short times.

\acknowledgments
AI is grateful to L. Peliti for long and interesting discussions.


\begin{thebibliography}{}

\bibitem{libchaber} {A. Simon, A. Libchaber}, Phys. Rev. Lett.,
68, 3375 (1992).

\bibitem{Benzi} {R.Benzi,G. Parisi, A. Sutera,A. Vulpiani}, SIAM
J. Appl. Math., 43, 565 (1983).



\bibitem{gammaitoni95}
{L. Gammaitoni, F. Marchesoni and S. Santucci.}
{Phys. Rev. Lett.} {\bf 74} (7), 1052-1055 (1995).

\bibitem{babic04}
{D. Babic, C. Schmitt, I. Poberaj and C. Bechinger.}
{Europhys. Lett.} {\bf 67} (2), p. 158 (2004). 

\bibitem{iwai01}
{T. Iwai.}
{Physica A} {\bf 300}, pp. 350-358 (2001)

\bibitem{dan05}
{D. Dan and A. M. Jayannavar.}
{Physica A} {\bf 345}, pp. 404-410 (2005)

\bibitem{evans}
D.~J.\ Evans, E.~G.~D.\ Cohen, and G.~P.\ Morriss, {\em Phys.\
Rev.\ Lett.} \textbf{71}, 2401 (1993); D.~J.\ Evans and D.~J.\
Searles, {\em Phys.\ Rev.\ E} \textbf{50}, 1645 (1994). 



\bibitem{gallavotti95}
{G. Gallavotti, E. G. D. Cohen.} {Phys. Rev. Lett.} {\bf 74}, pp.
2694 - 2697 (1995). 

\bibitem{kurchan98}
{Jorge Kurchan.} {J. Phys. A: Math. Gen.} {\bf 31}, pp. 3719?3729
(1998). J.~L.\ Lebowitz and H.~Spohn, {\em J.\ Stat.\ Phys.}
\textbf{95}, 333 (1999). 

\bibitem{ritort06}
{F. Ritort}
{J. Phys.: Condens. Matter } {\bf 18}, R531 (2006). 

\bibitem{Cohen}
R. van Zon and E.G.D. Cohen, {\em Phys. Rev. Lett.} {\bf 91} (11)
110601 (2003); {\em Phys. Rev. E} {\bf 67} 046102 (2003); {\em
Phys. Rev. E} {\bf 69} 056121 (2004).
R. van Zon, S. Ciliberto, E.G.D. Cohen, {\em Phys. Rev. Lett.}
{\bf 92} (13) 130601 (2004).



\bibitem{wang02}
{G.M. Wang, E.M. Sevick, E. Mittag, D. J. Searles and D. J.
Evans.}
{Phys. Rev. Lett.} {\bf 89}, 050601 (2002)

\bibitem{blickle06}
{V. Blickle, T. Speck, L. Helden, U. Seifert and C. Bechinger.}
{Phys. Rev. Lett.} {\bf 96}, 070603 (2006). 



\bibitem{joubaud07}
{F. Douarche, S. Joubaud, N. B. Garnier, A. Petrosyan, and S.
Ciliberto}
{Phys. Rev. Lett.} {\bf 97} (14), 140603 (2006). 

\bibitem{garnier}
N. Garnier, S. Ciliberto, {\em Phys.\ Rev.\ E} {\bf 71} 060101(R)
(2005). 


\bibitem{schuler05}
{S. Schuler, T. Speck, C. Tietz, J. Wrachtrup and U. Seifert.}
{Phys. Rev. Lett.} {\bf 94}, 180602 (2005). 


\bibitem{seifert07} T. Speck, V. Blickle, C. Bechinger and U.
Seifert, {Europhys.Lett}, {\bf 79}, 30002 (2007). 

\bibitem{saikia07}
{S. Saikia, R. Roy and A. M. Jayannavar.}
Phys. Lett. A, {\bf 369} pp. 367Ð371 (2007).

\bibitem{sahoo07}
{M. Sahoo, S. Saikia, M. C. Mahato and A. M. Jayannavar}
in press Physica A (2008). 

\bibitem{sekimoto96}
{K. Sekimoto} {J. Phys. Soc. Jpn.}
{\bf 66} (5), pp. 1234-1237 (1997). 

\bibitem{schmitt06}
{C. Schmitt, B. Dybiec, P. H\"anggi and C. Bechinger.}
{Europhys. Lett.} {\bf 74} (6), p. 937 (2006). 

\bibitem{seifert07b}
{U. Seifert}
{Eur. Phys. J. B}, {\bf 64} 3-4, pp. 423-431 (2008).

\bibitem{seifert05} U. Seifert {Phys. Rev. Lett.} {\bf 95}, 040602 (2005).

\bibitem{Taniguchi}T. Taniguchi and E. G. D. Cohen {J. Stat. Phys.} {\bf 126} 1 (2007)

\bibitem{mai07}
{T. Mai and A. Dhar.}
{Phys. Rev. E} {\bf 75}, 061101 (2007)


\bibitem{jop} P. Jop, A. Petrossyan, S. Ciliberto, {Europhys. Lett.}, {\bf 81} 50005 (2008).


\bibitem{al_pre07} A. Imparato, L. Peliti, G. Pesce, G. Rusciano, A. Sasso, {\it Phys. Rev. E}, {\bf 76} 050101R (2007).
\bibitem{SpSe} T. Speck and U. Seifert, J. Phys. A 38, L581 (2005).
\bibitem{al_w} A. Imparato and L. Peliti, Phys. Rev. E 72, 046114 (2005); A. Imparato and L. Peliti, Europhys. Lett. 70, 740  (2005).


\bibitem{mat_book} S. Wolfram, The Mathematica Book (Cambridge University Press, Cambridge, 1999).

\end{thebibliography}
\end{document}